# Magnetic Phase Diagrams of Multiferroic Hexagonal $RMnO_3$ (R=Er, Yb, Tm, and Ho)

F. Yen[1], C. dela Cruz[1], B. Lorenz[1], E. Galstyan[1], Y. Y. Sun[1], M. Gospodinov[2], and C.W, Chu[1,3,4]

[1] *Department of Physics and TcSUH, University of Houston, Houston ,TX 77204-5002*
[2] *Institute of Solid State Physics, Bulgarian Academy of Sciences, 1784 Sofia, Bulgaria*
[3] *Lawrence Berkeley National Laboratory,1 Cyclotron Road,Berkeley,CA 94720*
[4] *Hong Kong University of Science and Technology, Hong Kong, China*

## Abstract:

The magnetic phase diagrams of $RMnO_3$ (R = Er, Yb, Tm, Ho) are investigated up to 14 Tesla via magnetic and dielectric measurements. The stability range of the AFM order below the Néel temperature of the studied $RMnO_3$ extends to far higher magnetic fields than previously assumed. Magnetic irreversibility indicating the presence of a spontaneous magnetic moment is found near 50 K for R=Er, Yb, and Tm. At very low temperatures and low magnetic fields the phase boundary defined by the ordering of the rare earth moments is resolved. The sizable dielectric anomalies observed along all phase boundaries are evidence for strong spin-lattice coupling in the hexagonal $RMnO_3$. In $HoMnO_3$ the strong magnetoelastic distortions are investigated in more detail via magnetostriction experiments up to 14 Tesla. The results are discussed based on existing data on magnetic symmetries and the interactions between the Mn-spins, the rare earth moments, and the lattice.

## Introduction

Multiferroic hexagonal rare earth manganites have attracted special attention because of the coexistence of ferroelectric (FE) and magnetic orders. A phenomena resulting from the coupling and mutual interference of magnetism and ferroelectricity is a large magnetoelectric (ME) effect. This effect allows the magnetic field to tap into and control the dielectrical properties of the system [1] and vice versa with the electric field allowed to control the magnetic properties of the system [2]. Information may thus be stored both in the magnetic and electrical polarization instead of just the magnetic polarization and it may also be retrieved by sensing the magnetic moment or the ferroelectric polarization. A thorough understanding of the underlying physics between the coupling of these two orders may thus yield the possibility of developing novel devices that will improve memory storage densities.

For the hexagonal $RMnO_3$ (R = Sc, Y, In, Ho – Lu), there are two main structural properties that are essential for the multiferroicity. One is the ferroelectric order that sets in at high temperatures and results in ionic displacements breaking the inversion symmetry of the lattice [3]. The other feature is the presence of geometrical frustration caused by the antiferromagnetic (AFM) ordering of the Mn spins within a planar triangular lattice with ordering temperatures ranging from 70 – 130 K [4]. The linear ME effect in the hexagonal $RMnO_3$ is not allowed by symmetry (in zero magnetic field) [2]. However, the coupling of the two orders, namely the c-axis polarization and the in plane staggered AFM magnetization, is evident from strong dielectric anomalies at the magnetic phase transitions as observed for example in $HoMnO_3$ [5]. Therefore, it has been proposed that strong spin-lattice interactions do exist in all hexagonal $RMnO_3$ compounds [5, 6]. The existence of strong magneto-elastic effects has been shown through thermal conductivity [7] and thermal expansion experiments [8]. Furthermore, the complexity of the $HoMnO_3$ phase diagram at low temperatures [5, 6, 9, 10] adds more motivation in trying to understand the correlation between the magnetic orders and the dielectric properties in the various phases. Thus, the investigation of the phase diagram and the magnetic and dielectric anomalies at the phase boundaries is essential to detect the magneto-dielectric couplings and to associate them with the known magnetic structures and symmetries. The phase diagram of $HoMnO_3$ investigated recently up to 7 Tesla is extended to 14 Tesla to explore further the stability of the Néel phase boundary as well as the new low temperature phases discovered previously [9]. In addition, other hexagonal $RMnO_3$ with its rare earth ions having a magnetic moment (R=Er, Yb, and Tm) are studied for comparison. The phase boundaries are mainly monitored through the measurements of DC magnetization and the dielectric constant, ε. Heat capacity and AC magnetization measurements were performed at and close to zero magnetic fields.

## Experimental

An initial solid state synthesis is done with stoichiometric amounts of $Ho_2O_3$, $Tm_2O_3$ (99.9%), $Er_2O_3$, (99.9%) $Yb_2O_3$ (99.9%) and $MnO_2$ (99.995%) and annealed at 1100 $^{o}$C in an oxygen atmosphere for 24 hours. $HoMnO_3$, $TmMnO_3$, $ErMnO_3$ and $YbMnO_3$ single crystals were obtained by the high-temperature solution growth method in crucibles of pure platinum. Solvents of $PbO$:$PbF_2$:$B_2O_3$ (99.999%) were used as flux with a ratio of 0.5:0.45:0.05. The ratio of $RMnO_3$ powder and flux varies from 1:2.5 to 1:4. The temperature is increased at a rate of 50 $^{o}$C/h to 1300 $^{o}$C and maintained for 48 hours and subsequently lowered at a rate of 0.5 $^{o}$C/h to 920 $^{o}$C. At this temperature the crucible is taken out and the solvent is poured out. The obtained crystals remain on the ground and the walls of the crucible.

Silver or gold were sputtered onto the samples' surfaces perpendicular to the c-axis and used as parallel capacitor plates. The c-axis dielectric constant was measured by the high precision capacitance bridge AH 2500A (Andeen Hagerling) at 1 kHz. The Quantum Design's Physical Property Measurement System (PPMS) was used for temperature control to 1.8 K under external magnetic fields of up to 7 Tesla. High field measurements up to 14 Tesla were conducted in a superconducting magnet from Oxford Cryosystems with the $^{4}$He Variable Temperature Insert (VTI). The lowest temperature achieved in the VTI was 1.4 K. The specific heat data was obtained from the PPMS and the DC and AC magnetization were measured using the Quantum Design Magnetic Property Measuring System (MPMS). The external magnetic field was always applied along the c-axis.

The longitudinal magnetostriction was measured by a home made high precision capacitance dilatometer that was adapted to the Oxford VTI and 14 Tesla magnet.

## Results & Discussion

### $ErMnO_3$

At zero magnetic field $ErMnO_3$ becomes ferroelectric at $T_C$=835 K [3] and it orders antiferromagnetically at $T_N$=81 K [4] with a frustrated Mn spin arrangement corresponding to the $P\underline{6}_3\underline{c}m$ magnetic symmetry [11, 12]. The phase transition is clearly identified by a change in slope in the dielectric and magnetic data and a peak anomaly in the specific heat data (Fig. 1) in agreement with previous reports [13, 14]. The entropy associated to $T_N$ is 1.97 J/mole-K, 15% of the expected theoretical R*ln(2S+1) value (S=2). This value is consistent to what was obtained for the $HoMnO_3$ case [6]. Short range correlations at temperatures far higher than $T_N$ start to develop enhancing the heat capacity well above $T_N$ [15]. Thereby, only part of the expected change in entropy is detected at $T_N$. The strong correlations are also reflected in a negative thermal expansivity of the c-axis as is shown in the thermal expansion data for $HoMnO_3$ [8] and for $ErMnO_3$ [16] as well as in neutron diffraction data for $YMnO_3$ [17].

The inverse molar magnetic susceptibility along the c-axis and ab-plane behaves linearly at temperatures above 100 K. The Curie-Weiss temperature, $\theta_{CW}$, along the c-axis and ab-plane is extracted to be –75 K and –4 K, respectively. The effective moment for both interplane and intraplane magnetizations are $\mu_{eff}$=9.85 $\mu_B$, 92% of the expected theoretical value (10.69 $\mu_B$) and comparable to the 10.00 $\mu_B$ reported in [14]. The difference in $\theta_{CW}$ along the c-axis and ab-plane is similar to $HoMnO_3$ [6] and it confirms the strongly anisotropic character of the magnetic interactions with the Mn-spins confined to the ab-plane and the R-moments oriented along the c-axis [18].

Below $T_{RE}$=2.5 K the sharp increase of the heat capacity as well as the AC magnetic susceptibility indicates the onset of the ordering of the Er moments (Fig. 2). This phase is restricted only to extremely low magnetic fields. The derived phase boundary (Fig. 2c) shows that 0.05 Tesla is already enough to suppress this phase at 1.85 K (Fig. 3).

A third characteristic temperature, $T_{irr}$=55 K, is observed in the magnetic data. $T_{irr}$ is weakly dependent on the magnetic field up to 5 Tesla (the current limit of the magnetometer) and is defined by a small anomaly in the M(T) data, clearly visible in the derivative dM/dT shown in the insets of Figs. 4c and d. Along with $T_{irr}$, a splitting of the zero field cooling (ZFC) and field cooling (FC) DC magnetization data is also present (Fig. 4). This anomaly is present in many reports and will be discussed in a subsequent section.

The field dependence of $T_N$ is minute up to 7 T; it decreases only slightly from 81.5 K to 78 K (Fig. 5). At higher magnetic field, however, $T_N$ decreases faster and above 12.5 T ($T_N$=64 K) the AFM $P\underline{6}_3\underline{c}m$ phase is no longer stable. Instead, with decreasing temperature, the phase boundary traces back towards low fields. Upon decreasing temperature, at a constant magnetic field, there are two subsequent phase transitions associated with changes of the magnetic symmetry. In addition to distinct changes of the magnetization the phase boundary is clearly defined by sharp anomalies of the measured dielectric constant (Figs. 6, 7). The critical temperatures derived from magnetic and dielectric data are consistent with each other. The phase diagram of Fig. 5 summarizes data derived from dc and ac magnetic susceptibility as well as dielectric measurements conducted upon increasing and decreasing magnetic field and temperature. It is worth noting that no thermal or field hysteresis could be detected in crossing the transition temperature.

The shape of the $T_N$ phase boundary and the reentrant behavior in our magnetic phase diagram is similar to that reported by Fiebig et al, [11, 12]. However, the stability range of the AFM $P\underline{6}_3\underline{c}m$ phase of our crystal extends far beyond the 4 Tesla limit of Ref. [11]. The phase boundary for the $P\underline{6}_3\underline{c}m$ phase is a second order phase transition as it has been proven by a recent study of the critical behavior of $YMnO_3$ at $T_N$=73 K [19]. Our results show no thermal or magnetic hysteresis upon crossing the Néel phase boundary within the phase diagram of Fig. 5. This is in contrast to the results derived from second harmonic generation measurements that have been interpreted in terms of a wide hysteretic area existing in the phase diagrams of $ErMnO_3$, $TmMnO_3$, and $YbMnO_3$ [11, 12]. The stability of the $P\underline{6}_3\underline{c}m$ phase towards much higher fields is consistent with the $HoMnO_3$ magnetic phase diagram where the $P\underline{6}_3\underline{c}m$ magnetic structure exists even beyond 14 Tesla (see phase diagram of $HoMnO_3$ below). At temperatures below $T_{RE}$=2.5 K the magnetic order of the Er moments evolves as is clearly seen in the anomalies of the c-axis magnetic susceptibility and the specific heat (Fig. 2). The M(H) measurements at low temperatures do not show any signature of field hysteresis that might indicate a sizable ferromagnetic (FM) moment, in contrast to previous reports by Sugie et al. [20] and Fiebig et al. [11] who attributed the field hysteresis to the FM order of the Er moment. The magnetization data shown in Fig. 3 include both, data taken with increasing and decreasing magnetic field and the maximum field applied was 2 Tesla. Although our low-field magnetic measurements revealed a minute irreversible magnetic behavior below 50 K (Fig. 4c) the spontaneous moment involved must be small and could arise in domain boundaries of the primarily AFM magnetic structure. Differences in sample quality or composition could be the origin of the observed discrepancies (samples investigated in Ref. [11] and [20] are apparently from one and the same source). Deviations from the ideal stoichiometry, for example, can lead to the presence of Mn ions with different valence than 3+ giving rise to a weak FM or canted AFM moment at low T. The physical origin of the discrepancies as discussed above still needs further investigations.

## $YbMnO_3$

The magnetic and dielectric measurements for $YbMnO_3$ reveal similar $\varepsilon(T)$ and $\chi(T)$ anomalies at the Néel phase boundary like $ErMnO_3$. At very low temperatures, the ytterbium moment order sets in at $T_{RE}$=4 K and this phase is quickly suppressed by external magnetic fields. At 2 K a field of 0.07 Tesla does destroy the order of Yb moments (Fig. 8). The stability region of the AFM $P\underline{6}_3\underline{c}m$ phase extends beyond 14 Tesla, the current field limit of the experiment. However, the reentrant behavior is clearly seen and the Néel temperature approaches zero at a critical field of 3.2 Tesla. This value of the critical field is in good agreement with the metamagnetic transition reported in Ref. [20] in the M(H) data at 1.8 K. At the lowest temperature (1.8 K) both phase transitions at $T_{RE}$ and at $T_N$ are well defined by sharp anomalies of the magnetization as a function of the magnetic field (Fig. 9). Similar to $ErMnO_3$, no thermal or magnetic field hysteresis has been observed across both magnetic phase boundaries. The phase boundary reported in Ref. [20] for temperatures below 10 K is in good agreement with our results (Fig. 8). In addition, we extend the phase diagram to 14 Tesla which allows us to extrapolate the stability region of the $P\underline{6}_3\underline{c}m$ phase. The low temperature phase boundary of the $P\underline{6}_3\underline{c}m$ phase appears to merge with $T_N$ at a field slightly larger than 14 Tesla.

Comparing our results with the phase diagram derived from second harmonic generation optical experiments a similar discrepancy as discussed for $ErMnO_3$ above becomes obvious. Whereas the stability region of the $P\underline{6}_3\underline{c}m$ phase of Ref. [12] is limited to fields below 5 Tesla and a wide area of hysteresis occupying the major part of the T-H phase diagram has been reported, our measurements show no indication of hysteretic behavior across the phase boundaries and the $P\underline{6}_3\underline{c}m$ phase extends beyond the 14 Tesla limit of our investigation. The zero temperature transition revealed in this study at about 3 Tesla is in excellent agreement with the magnetic data at 1.8 K of Ref. [20] but it is clearly larger than the transition field of less than 2 Tesla derived in [12]. Reconciliation of this conflict seems difficult if one assumes that the samples of Ref. [12] and [20] are from the same source. The low temperature phase characterized by the Yb moment order is stable only at low fields. The zero field value of $T_{RE}$ is comparable with the value derived from magnetic measurements by Sugie et al. [20].

**$TmMnO_3$**

The magnetic phase diagram of $TmMnO_3$ (Fig. 10) is similar to the phase diagrams of $ErMnO_3$ and $YbMnO_3$ with the exception that the ordering of the rare earth magnetic moment could not be observed from temperatures above 1.8 K. The magnetic phase transitions are nicely visible in distinct and sharp anomalies of the dielectric constant, shown in Fig. 11. At zero magnetic field $T_N$=84 K and the maximum field below which the AFM phase is stable is limited to 10.5 Tesla. The phase boundary of the $P\underline{6}_3\underline{c}m$ phase extrapolates to zero at the critical field of 4.3 Tesla. Between this field and the maximum field of 10.5 Tesla two phase transitions are observed upon decreasing temperature (Fig. 11). The larger stability range of the $P\underline{6}_3\underline{c}m$ phase and the absence of thermal or magnetic field hysteresis are similar features as observed in $ErMnO_3$ and $YbMnO_3$, but they are, again, in striking contrast to the phase diagram reported earlier [12].

It is remarkable that no sharp phase change was detected in $ErMnO_3$ and $TmMnO_3$ upon decreasing temperature at magnetic fields above the critical field of the $P\underline{6_3}\underline{c}m$ phase. The same probably applies to the phase diagram of $YbMnO_3$ although the field range of our measurements does not exceed the critical field for this compound (> 14 T). It appears unlikely that our highly sensitive dielectric measurements could not reveal a sharp phase boundary between the high-temperature phase and the low-T $P6_3\underline{cm}$ magnetic structure. However, upon decreasing temperature the transition to the $P6_3\underline{cm}$ at high magnetic fields can be smooth as characteristic for thermodynamic changes above a critical point. The maximum field of the $P\underline{6_3}\underline{c}m$ phase would then represent a critical point in the T-H phase diagram. A similar situation was observed in $Ni_3V_2O_8$ and discussed in Ref. [21]. In this compound a smooth transition from the high-T (PM) phase to a low-T (AFM) phase was detected for applied fields (along the c-axis of $Ni_3V_2O_8$) higher than a critical one without crossing a sharp phase boundary. The same physical mechanisms may explain our observations in the hexagonal $RMnO_3$ compounds. It implies that the high-temperature paramagnetic phase in magnetic fields exhibits the same symmetry as the low-T $P6_3\underline{cm}$ phase and that the field-induced canting of the Mn-spins towards the c-axis also causes a staggered in-plane spin modulation according to the $P6_3\underline{cm}$ symmetry. The antisymmetric Dzyaloshinskii- Moriya interaction can generate a bilinear coupling between the c-axis and in-plane magnetic moments resulting in the particular magnetic order. Since the low-field $P\underline{6_3}\underline{c}m$ magnetic structure does not allow for a c-axis moment a sharp transition between this phase and the high-field $P6_3\underline{cm}$ phase is immanent, as observed in our experiments.

**$HoMnO_3$**

The $HoMnO_3$ phase diagram is far more complex than the ones for $ErMnO_3$, $TmMnO_3$ and $YbMnO_3$ as it has been shown from many previous works [6, 9, 10, 22] in a magnetic field range up to 7 Tesla. $HoMnO_3$ exhibits novel phases with at least 5 of them characterized by sharp dielectric and magnetic anomalies in the low temperature region [6, 9]. In addition there is an intermediate phase (INT) in the H-T phase diagram that is believed to be a phase in which the spins of the manganese are at an angle with the hexagonal a-axis in between the two "locked-in" positions given by the $P\underline{6_3}\underline{c}m$ magnetic structure (HT1 phase, stable for 33 K < T < $T_N$ at zero field, Mn-spins perpendicular to the a-axis) and the $P\underline{6_3}c\underline{m}$ configuration (HT2 phase for 5.2 K < T < 33 K, Mn-spins parallel to the hexagonal a-axis) [5]. The magnetic structure of this INT phase is described by the $P\underline{6_3}$ symmetry group and has lower magnetic anisotropy. Extending the phase diagram of $HoMnO_3$ to higher fields (up to 14 Tesla), it is found that the $P\underline{6_3}\underline{c}m$-HT1 phase is far more stable as compared to the other hexagonal $RMnO_3$ with the Néel temperature decreasing only by 1 K at 13 Tesla. Unlike for R=Er, Tm, and Yb (Figs. 4, 8, 10) $T_N$ cannot be traced to zero temperature in the high-field phase diagram shown in Fig. 12. Instead, we show that the low-temperature phase boundary denoted by $T_4$ in our previous work [9] (see also Figs. 12, 14) decreases towards higher field and $T_4$ approaches zero at the critical field of 7.7 Tesla (measured at 1.4 K). No further anomaly could be detected in our dielectric measurements above that critical field and at low temperatures. Assuming that the $P\underline{6_3}\underline{c}m$ phase of $HoMnO_3$ becomes unstable at high enough magnetic fields (as in the case of $ErMnO_3$, $YbMnO_3$, and $TmMnO_3$) this critical field must be considerably larger than the 14 Tesla limit of the current experiment.

It is remarkable that in all hexagonal $RMnO_3$ the magnetic phase transitions are associated with sharp and distinct anomalies of the dielectric constant, as e.g. shown in Figs. 6, 7, and 11. This is unambiguous evidence for the existence of strong spin-lattice interactions resulting in the magneto-dielectric coupling observed particularly in crossing the various magnetic phase boundaries. The onset or change of the magnetic orders causes the lattice to respond via magnetoelastic coupling. This effect was revealed in the most convincing manner by detecting the macroscopic lattice strain of $HoMnO_3$ in thermal expansion measurements [8]. However, in the complex H-T phase diagram of $HoMnO_3$ it is not clear whether magnetoelastic effects play a leading role in all phases and how external magnetic fields couple to the lattice. We have therefore investigated the magnetostriction of $HoMnO_3$ along the c-axis at different constant temperatures and compared the results with similar isothermal measurements of the dielectric constant.

Fig. (13) shows the relative change of the c-axis length with the external field at higher temperatures where the transitions from HT2 ($P\underline{6}_3\underline{c}m$) to INT ($P\underline{6}_3$) to HT1 ($P\underline{6}_3c\underline{m}$) phases take place with increasing H. Data for 15 K and 8 K are shown and the latter curve is offset by 5 units for better clarity. The three phases are labeled for the 15 K data set and the phase transitions are indicated by dotted lines at $H_2$ and $H_1$. The c-axis strain in the intermediate $P\underline{6}_3$ phase shows a far stronger response to the external field than in the $P\underline{6}_3c\underline{m}$ and $P\underline{6}_3\underline{c}m$ phases. The magnetostrictive coefficient (the field derivative of c(H)/c(0)) shows a plateau-like enhancement between $H_2$ and $H_1$. This result is in excellent agreement with similar plateau-like enhancements of the dielectric constant [11] and the ac magnetic susceptibility [6] reported and discussed by us earlier. All the results show that the $P\underline{6}_3$ INT phase is a very "soft" phase, the physical properties of which are extremely susceptible to external fields such as the electric or magnetic field. This softness is possibly due to the lower magnetic anisotropy in this phase. The similar behavior of the dielectric constant, the ac susceptibility, and the lattice strain reveal the correlation between the properties and the important role of the magetoelastic couplings. The softness of this phase can be understood as a consequence of the particular lower magnetic symmetry. In the neighboring $P\underline{6}_3\underline{c}m$ and $P\underline{6}_3c\underline{m}$ phases the Mn-spin angle in the hexagonal plane is locked at two stable positions with respect to the a-axis, 90˚ and 0˚, respectively. In the $P\underline{6}_3$ phase, however, the angle is intermediate between the stable positions and it can even be a function of temperature or field in the phase diagram or it can be easily affected by external fields. This leads to a larger response of the magnetic, dielectric, and structural properties to external fields and the observed enhancements of the response functions as discussed above.

Below 5 K the phase diagram of $HoMnO_3$ is most complex and various phases and transitions have been discussed [6, 9, 10] (Fig. 14). To add new insight into this interesting region of the phase diagram and into the role of magnetoelastic effects the lattice strain along the c-axis has been measured and compared to the magnetic and dielectric data. Of particular interest is a narrow region close to 2 Tesla and stretching from 3 to about 6 K (dashed line in Fig. 14) that exhibits the sharp peak of the dielectric constant and the ac magnetic susceptibility. Furthermore, the dielectric constant indicates the different phase boundaries by distinct, step-like changes. At the lowest temperatures two dome shaped phases, LT1 ($0.5\ T < H < 2\ T$) and LT2 ($2\ T < H < 7.7\ T$) determine the landscape in the H-T phase diagram. The results of the isothermal magnetostrictive measurements, shown in Fig. 15, clearly reveal the c-axis anomalies at most of the magnetic phase transitions. With increasing field and for temperatures below 3.3 K the entrance into the LT1 phase is associated with a sharp increase of the c-axis length. This anomaly disappears at higher temperature limiting the stability of this phase to $T<3.3$ K, as discussed earlier [6]. A second sharp increase of the c-axis is observed at about 2 Tesla at the lowest temperature (1.43 K) of this investigation. This c-axis anomaly develops into two separate steps at higher temperature indicating two subsequent phase transitions in the field range close to 2 Tesla. It is exactly in this field-temperature range where a sharp peak of the dielectric constant (Fig. 16) and of the ac magnetic susceptibility have been reported [6, 9], before the transition from the $P\underline{6}_3$ intermediate to the $P\underline{6}_3\underline{c}m$ HT1 phase takes place. The separation of the two lattice anomalies in this narrow field range is clearly seen in the derivative of c(H) with respect to H showing two well separated, distinct peaks close to 2 Tesla (inset of Fig. 15 for T=3 K). Therefore, we associate the first increase of c with the peak of the dielectric constant and the second step with the transition $P\underline{6}_3 \rightarrow P\underline{6}_3\underline{c}m$. At higher temperatures the first anomaly gradually decreases as the peaks of the dielectric constant and the ac magnetic susceptibility become broader. However, the transition into the $P\underline{6}_3\underline{c}m$-phase remains characterized by a sharp change of slope of c(H) at the critical field $H_1$ (Figs. 13 and 15) for all temperatures below the spin reorientation phase transition, $T_{SR}$=33 K.

The distinct and abrupt anomalies in crossing the various phase boundaries detected in the thermal expansion [8] and in the current magnetostriction data prove the strong correlation of magnetic order, the dielectric properties, and the lattice strain. However, no anomalous change of the c-axis length could be detected along the phase boundary between the HT1 and LT2 phases although the dielectric constant shows a sudden drop in crossing this phase boundary into the LT2 phase (stable between 2 Tesla and 7.7 Tesla for temperatures below 4.6 K). Isothermal measurements of $\varepsilon(H)$ at 1.4 K and 4.24 K are shown in Fig. 16 and the drop of $\varepsilon$ at the phase boundary $H_4$ is indicated by the dotted arrows. A similarly sharp decrease of $\varepsilon$ can also be detected upon decreasing temperature at constant magnetic field between 2 Tesla and 7.7 Tesla. This is the only high-field phase boundary in the H-T phase diagram that shows a sizable step of the dielectric constant but no anomaly in the c-axis lattice constant. This is an indication that the coupling of the magnetic order with the dielectric properties does not involve a sizable lattice distortion along this phase boundary.

At low magnetic fields the c-axis increases nonlinearly with the magnetic field. This is characteristic of the quadratic magnetoelastic coupling between the AFM magnetic moments and the lattice strain. In the high-field phases (HT1, LT2), however, a linear magnetoelastic effect is clearly observed over a wide field range, c(H) increases linearly with H, as shown in Fig. 15. This linear magnetoelastic coupling is apparently induced by the magnetic field and it is associated with the phase transition into the HT1 and LT2 phases, similar to the linear magnetoelastic effect observed in multiferroic $TbMnO_3$ [23]. The origin of the linear magnetoelastic coupling along the c-axis is not clear yet but it appears conceivable that it is associated with the alignment of the Ho moments with the c-axis magnetic field. Metamagnetic transitions have in fact been observed in low-temperature magnetization data with increasing field [6]. This raises the question about the involvement of the Ho moments, their magnetic exchange interactions, and their interactions with the frustrated spin order of the Mn ions. The complexity of the magnetic phase diagram at low temperatures is an expression of the strong correlations of the systems of Mn spins, Ho moments, the ferroelectric displacements, and their interactions with the lattice.

## Magnetic irreversibility in $RMnO_3$ at low fields and the coupling between Mn-spins and R-moments

Aside from $HoMnO_3$, magnetization measurements for $ErMnO_3$, $YbMnO_3$ and $TmMnO_3$ (Fig. 4) show subtle anomalies at temperatures near $T_{irr} \approx 50$ K. The FC and ZFC data are different below $T_{irr}$ indicating the irreversibility of the magnetic properties. This type of splitting is well known for ferromagnets and it is caused by the spontaneous magnetization and the response of the FM domains with respect to the external field depending on the cooling history (with or without applied field). In perfect antiferromagnets the uncompensated, spontaneous moments can only arise in the domain walls and the observed difference in FC and ZFC magnetization is therefore small. The onset of this irreversibility at a temperature around 50 K or lower needs a more careful discussion. For $HoMnO_3$ it was shown that close to this temperature, $T_{irr} \approx 33$ K, the manganese spins rotate collectively by 90° and part of the Ho moments start to order antiferromagnetically with the moments aligned with the c-axis [24, 25]. For R=Er, Tm, Yb the situation is less clear. $T_{irr}$ decreases modestly with increasing field and the splitting is always present. For example, for $ErMnO_3$, $T_{irr}$ decreases roughly by 1 K every 1 Tesla traceable up to 5 Tesla (the magnetometer's limit). The splitting is not observed when H is along the hexagonal ab-plane allowing the dismissal of $T_{irr}$ being caused by magnetic impurities. In compounds such as $ScMnO_3$, $LuMnO_3$ and $YMnO_3$, broad maxima of the thermal conductivity [7] and the excess magnetic specific heat have been reported at $T_{irr}\approx 50$ K [26]. Neutron powder diffraction data of $ScMnO_3$ show an additional magnetic transition at around 50 K [27] which should come from the $Mn^{3+}$ spins since $Sc^{3+}$, $Lu^{3+}$ and $Y^{3+}$ are non-magnetic. These anomalies have not been understood so far and it is not clear if they are associated with the magnetic irreversibility observed close to 50 K (Fig. 4).

$T_{irr}$ appears to be intrinsic in the hexagonal $RMnO_3$ including $HoMnO_3$ where FC and ZFC susceptibilities also split close to $T_{SR}$=33 K (Fig. 4a). The existence of magnetic irreversibility in the c-axis susceptibility indicates the presence of a spontaneous magnetic moment along the hexagonal c-axis. Fiebig et al. [12] have found phase coexistence in several $RMnO_3$ and magnetic hysteresis attributed to ferrimagnetic (at low fields) and ferromagnetic (high fields) orders of the Yb sublattice in $YbMnO_3$. However, the current magnetization data for our $YbMnO_3$ crystals (Fig. 9) do not show any sign of M(H) hysteresis in magnetic fields up to 5 Tesla. It should be noted that the data shown in Fig. 9 include measurements upon increasing and decreasing magnetic field and any possible M(H) hysteresis is below the resolution limit of the experiment. Alternatively, the spontaneous c-axis moment causing the FC-ZFC irreversibility could be found in the domain boundaries of the AFM domains defined by the Mn-spins. It is well known that in systems with d-electron magnetic order and f-electrons present the magnetic moments of the f-electrons can be polarized by the d-spin order. In the hexagonal $RMnO_3$ the two systems of magnetic moments are perpendicular, the Mn-d-spins are confined to the ab-plane and the R-f-moments are oriented along the c-axis. Therefore, the exchange coupling between the two magnetic systems is expected to be weak. The magnetic dipolar interaction, for example, can result in a partial polarization of the R-moments below the AFM ordering temperature of the Mn-spins. Considering the R-moments exposed to the internal field of the Mn-spins it can be shown that certain AFM magnetic structures of the Mn-spin system favor an AFM polarization of the R-moments [28]. Considering the formation of AFM domains in the Mn-spin structure the R-moments will follow the domain structure and spontaneous R-moments oriented along the c-axis will arise in the domain walls. These moments can explain the observed irreversibility in FC and ZFC magnetization measurements. At temperatures close to $T_N$ this effect is too small to be observed since the sublattice magnetization of the Mn-spin system is low and it increases only towards lower temperature. The coupling between the Mn-spins and the R-moments will grow stronger with decreasing temperature and FC and ZFC magnetization data should split at $T_{irr} < T_N$. The direct exchange between the R-moments is expected to become relevant at far lower temperatures only, typically below 10 K. The above proposal may explain the observed irreversible behavior of M(T), the small magnitude of the overall effect, and the similar value of $T_{irr}$ for R=Er, Yb, Tm.

In the case of $HoMnO_3$ the coupling between the d- and f- magnetic subsystems is apparently stronger, possibly because of the larger magnetic moment of the $Ho^{3+}$. This results in the major change of the magnetic order at $T_{SR}$ = 33 K with the Mn-spins rotating by 90° changing their order from $P\underline{6}_3\underline{c}m$ to $P\underline{6}_3c\underline{m}$ and the Ho-moments becoming partially ordered. The dipolar field due to the Mn-spins at the Ho-sites is stronger in the $P\underline{6}_3c\underline{m}$ phase (below $T_{SR}$) and the system can gain energy from the dipolar interaction between Mn-spins and Ho-moments in this phase [28]. This is what possibly triggers the Mn-spin rotation and the simultaneous onset of the partial Ho-moment order. At very low temperatures the exchange between the rare earth moments becomes essential and triggers the order of the R-moments in $ErMnO_3$ and $YbMnO_3$ at $T_{RE}$ = 2.5 K and 4.0 K, respectively. In $HoMnO_3$ a rapid increase of the Ho-moment near 5 K has been reported [24] and heat capacity [6], dielectric [9], and magnetic [6] data indicate a sharp phase transition into a low-temperature phase. Second harmonic generation optical measurements have identified the magnetic symmetry of the low-temperature phase as $P6_3cm$ [12]. The in-plane alignment of the Mn-spins in this phase is similar to the one of the $P\underline{6}_3\underline{c}m$ phase but with a different phase sequence along the c-axis. A more careful inspection of the 5 K transition shows that it is actually split into two subsequent transitions as indicated by two distinct anomalies of various physical quantities [6, 9, 29] with a temperature separation of only 0.3 K. It appears conceivable that the magnetic exchange between the Ho-moments triggers the first transition at 5.2 K resulting in the reported increase of the Ho sublattice magnetization and that the second transition at 4.9 K is the rotation of the Mn-spins triggered by the increase of the Ho-moment and their coupling to the Mn-spins. Among all hexagonal $RMnO_3$ the $HoMnO_3$ seems to be the only compound that shows this complex phase sequence with a multitude of sharp phase transitions and different phases in the T-H phase diagram. Whether this is related to the magnitude of the magnetic moment of the $Ho^{3+}$-ion or to other effects has yet to be investigated. The sequence of magnetic structures and phase transitions in other rare earth manganites is far less complex although coexistence of different phases has been reported for some compounds [12].

## Summary and Conclusions

The magnetic phase diagrams of $RMnO_3$ (R=Ho, Er, Yb, Tm) have been re-investigated. It was found that the stability of the $P\underline{6}_3\underline{c}m$ phase extends to far higher magnetic fields than previously assumed. The maximum fields for the $P\underline{6}_3\underline{c}m$ structure are 12.5 and 10.5 Tesla for R=Er and R=Tm, respectively. For R=Yb and R=Ho the maximum field is above 14 Tesla, the field limit of the current investigation. At low temperatures the AFM order of the Er and Yb moments is observed and the stability range of the R-ordered phases in the M-T phase diagram was determined. The rare earth moment order is completely suppressed by low magnetic fields of about 0.07 Tesla. No magnetic or thermal hysteresis effects could be observed in crossing various phase boundaries or in the different phases. Sharp and distinct anomalies of the dielectric constant are revealed at all magnetic phase boundaries indicating a strong coupling of the magnetic orders with the lattice.

Magnetic irreversibility in form of a small difference between field cooling and zero-field cooling magnetization data exists in all compounds below an irreversibility temperature of about 50 K (33 K in $HoMnO_3$). This effect was attributed to a spontaneous magnetic moment of the rare earth moment arising in the magnetic domain boundaries. It was interpreted as a signature of a magnetically polarizing effect the Mn-spin system exhibits on the rare earth moment.

The magnetoelastic effects were further investigated in the complex phase diagram of $HoMnO_3$ through magnetostriction measurements in fields up to 14 Tesla. Several phase transitions previously described could be associated with abrupt changes of the c-axis lattice parameter further lending support to the suggestion that strong spin-lattice interactions play a prominent role in the coupling of magnetic moments, dielectric properties, and the ionic displacements.

## Acknowledgements

This work is supported in part by the T.L.L. Temple Foundation, the J.J. and R. Moores Endowment, and the State of Texas through the TCSUH and at LBNL by the DOE. The work of M. M. G. is supported by the Bulgarian Science Fund, grant No. F 1207.

## References

[1] N. Hur, S. Park, P. A. Sharma, J. S. Ahn, S. Guha, and S-W. Cheong: Electric polarization reversal and memory in a multiferroic material induced by magnetic fields. Nature **429**, 392 (2004).

[2] T. Lottermoser, T. Lonkai, U. Amann, D. Hohlwein, J. Ihringer, and M. Fiebig: Magnetic phase control by an electric field. Nature **430**, 541 (2004).

[3] Ph. Coeuré, P. Guinet, J. C. Peuzin, G. Buisson, and E. F. Bertaut: Ferroelectric properties of hexagonal orthomanganites of yttrium and rare earths. Proceedings of the International Meeting on Ferroelectricity **1**, 332-340 (1966) Prague.

[4] W. C. Koehler, H. L. Yakel, E. O. Wollan, and J. W. Cable: A note on the magnetic structures of rare earth manganese oxides. Physics Letters **9**, 93-95 (1964).

[5] B. Lorenz, A. P. Litvinchuk, M. M. Gospodinov, and C. W. Chu: Field-Induced Reentrant Novel Phase and a Ferroelectric-Magnetic Order Coupling in $HoMnO_3$. Physical Review Letters **92**, 087204 (2004).

[6] B. Lorenz, F. Yen, M. M. Gospodinov, and C. W. Chu: Field-induced phases in $HoMnO_3$ at low temperatures. Physical Review B **71**, 014438 (2005).

[7] P. A. Sharma, J. S. Ahn, N. Hur, S. Park, Sung Baek Kim, Seongsu Lee, J.-G. Park, S. Guha, and S.-W. Cheong: Thermal Conductivity of Geometrically Frustrated, Ferroelectric $YMnO_3$: Extraordinary Spin-Phonon Interactions. Physical Review Letters **93**, 177202 (2004).

[8] C. dela Cruz, F. Yen, B. Lorenz, Y. Q. Wang, Y. Y. Sun, M. M. Gospodinov, and C. W. Chu: Strong spin-lattice coupling in multiferroic $HoMnO_3$: Thermal expansion anomalies and pressure effect. Physical Review B **71**, 060407(R) (2005).

[9] F. Yen, C. R. dela Cruz, B. Lorenz, Y. Y. Sun, Y. Q. Wang, M. M. Gospodinov, and C. W. Chu: Low temperature dielectric anomalies in HoMnO$_3$: The complex phase diagram. Physical Review B **71**, 180407(R) (2005).

[10] O. P. Vajk, M. Kenzelmann, J. W. Lynn, S. B. Kim, and S.-W. Cheong Magnetic Order and Spin Dynamics in Ferroelectric $HoMnO_3$. Phys. Rev. Letters **94**, 087601 (2005).

[11] M. Fiebig, C. Degenhardt, and R. V. Pisarev: Interaction of Frustrated Magnetic Sublattices in $ErMnO_3$. Phys. Rev. Letters **88**, 027203 (2002).

[12] M. Fiebig, Th. Lottermoser, and R. V. Pisarev: Spin-rotation phenomena and magnetic phase diagrams of hexagonal $RMnO_3$. J. Appl. Phys. **93**, 8194 (2003).

[13] N. Iwata and K. Kohn: Dielectric Anomalies at Magnetic Transitions of Hexagonal Rare Earth Manganese Oxides $RMnO_3$. J. Phys. Soc. Jpn. **67**, 3318 (1998).

[14] M. Chandra Sekhar, Seongsu Lee, Gwangho Choi, Changhee Lee, and J.-G. Park: Doping effects of hexagonal manganites $Er_{1-x}Y_xMnO_3$ with triangular spin structure. Phys. Rev. B **72**, 014402 (2005).

[15] J. Park, J.-G. Park, Gun San Jeon, Han-Yong Choi, Changhee Lee, W. Jo, R. Bewley, K. A. McEwen, and T. G. Perring: Magnetic ordering and spin-liquid state of $YMnO_3$. Phys. Rev. B **68**, 104426 (2003).

[16] J. Park, U. Kong, S. I. Choi, J.-G. Park, C. Lee, and W. Jo: Magnetic structure studies of $ErMnO_3$. Appl. Phys. A **74**, S802 (2002).
[17] J. Park, U. Kong, A. Pirogov, S. I. Choi, J.-G. Park, Y. N. Choi, C. Lee, and W. Jo: Neutron-diffraction studies of $YMnO_3$. Appl. Phys. A **74**, S796 (2002).
[18] T. J. Sato, S. –H. Lee, T. Katsufuji, M. Masaki, S. Park, J. R. D. Copley, and H. Takagi: Unconventional spin fluctuations in the hexagonal antiferromagnet $YMnO_3$. Phys. Rev. B **68**, 014432 (2003).
[19] M. Tachibana, J. Yamazaki, H. Kawaji, and T. Atake: Heat capacity and critical behavior of hexagonal $YMnO_3$. Phys. Rev. B **72**, 064434 (2005).
[20] H. Sugie, N. Iwata, and K. Kohn: Magnetic Ordering of Rare Earth Ions and Magnetic-Electric Interaction of Hexagonal $RMnO_3$ (R=Ho, Er, Yb or Lu). J. Phys. Soc. Jpn. **71**, 1558 (2002).
[21] M. Kenzelmann, A. B. Harris, A. Aharony, O. Entin-Wohlman, T. Yildirim, Q. Huang, S. Park, G. Lawes, C. Broholm, N. Rogado, R. J. Cava, K. H. Kim, G. Jorge, and A. P. Ramirez, Phys. Rev. B **74**, 014429 (2006).
[22] M. Fiebig, C. Degenhardt, and R. V. Pisarev: Magnetic phase diagram of $HoMnO_3$. J. Appl. Phys., **91**, 8867 (2002).
[23] N. Aliouane, D. N. Argyriou, J. Strempfer, I. Zegkinoglou, S. Landsgesell, and M. v. Zimmermann: Field-induced linear magnetoelastic coupling in multiferroic $TbMnO_3$. Phys. Rev. B **73**, 020102(R) (2006).
[24] Th. Lonkai, D. Hohlwein, J. Ihringer, and W. Prandl: The magnetic structures of $YMnO_3$ and $HoMnO_3$. Appl. Phys. A **74**, S843 (2002).
[25] A. Muñoz, J. A. Alonso, M. J. Martínez-Lope, M. T. Casáis, J. L. Martínez, and M. T. Fernández-Díaz: Evolution of the Magnetic Structure of Hexagonal $HoMnO_3$ from Neutron Powder Diffraction Data. Chem. Materials **13**, 1497 (2001).
[26] D.G. Tomuta, S. Ramakrishnan, G. J. Nieuwenhuys, and J. A. Mydosh: The magnetic susceptibility, specific heat and dielectric constant of hexagonal $YMnO_3$, $LuMnO_3$ and $ScMnO_3$. J. Phys: Condens. Mat. **13**, 4543 (2001).
[27] A. Muñoz, J. A. Alonso, M. J. Martínez-Lope, M. T. Casáis, J. L. Martínez, and M. T. Fernández-Díaz: Magnetic structure of hexagonal $RMnO_3$ (R=Y, Sc): Thermal evolution from neutron powder diffraction data. Phys. Rev.B **62**, 9498 (2000).
[28] B. Lorenz, unpublished.
[29] E. Galstyan, unpublished.

**Captions:**

Figure 1:
a) Magnetic susceptibility along the c-axis for $ErMnO_3$. b) Dielectric constant curve along the c-axis at 1 kHz. c) Heat capacity curve at H=0.

Figure 2:
a) Heat capacity data at low temperature for $ErMnO_3$, the sharp step at $T_{RE}$=2.5 K indicates the ordering of $Er^{3+}$. b) Field dependence of the real part of the ac magnetic susceptibility at low temperatures. c) Phase boundary of the low temperature phase.

Figure 3:
M(H) of $ErMnO_3$ in both increasing and decreasing H. The change of slope defines the phase boundary to the low temperature phase with AFM $Er^{3+}$ moment order. No hysteresis is observed in the M(H) data.

Figure 4:
Magnetic irreversibility in $RMnO_3$. Open (closed) symbols indicate FC (ZFC) M(T) scans for a) $HoMnO_3$ at 8 Oe, b) $TmMnO_3$ at 1 Tesla, c) $ErMnO_3$ at 1 Tesla, and d) $YbMnO_3$ at 1 Tesla. The field is aligned with the c-axis. The inset to Figs. 4c,d show the derivative, dM/dT, revealing a subtle anomaly at $T_{irr}$. This anomaly is too small for $TmMnO_3$ to be resolved in Fig. 4b.

Figure 5:
Magnetic phase diagram of $ErMnO_3$. The phase boundary is derived from dc magnetization, dielectric constant, and ac magnetic susceptibility measurements upon in creasing and decreasing temperature and magnetic fields. The magnetic symmetries listed in different regions of the phase diagrams of this figure and of Figs. 8, 10, 12, 14 are those revealed in second harmonic generation measurements [11, 12].

Figure 6:
Dielectric constant (a) and magnetization (b) as function of magnetic fields. Sharp anomalies of both quantities define the phase boundaries. The dotted line in (a) marks the dielectric anomalies at $T_N$.

Figure 7:
High field ε(T) of $ErMnO_3$. Two sharp anomalies (indicated by arrows) show the two subsequent phase transitions in crossing through the $P\underline{6}_3\underline{c}m$ phase. Above 12.5 Tesla this phase is no longer stable. Different ε(T) curves are vertically offset for clarity. H is applied along the c-axis.

Figure 8:
Magnetic Phase Diagram of $YbMnO_3$. The inset shows the low-field low-temperature phase boundary defining the region of $Yb^{3+}$ moment order.

Figure 9:

M(H) measurements for $YbMnO_3$ showing the evolution of both of the phase transitions at different temperatures. Increasing and decreasing field for the 2 K curve is shown. For all other curves, data have been measured in both field directions but only data for increasing field is shown in the figure for clarity. No hysteresis could be observed in all data sets indicating the absence of FM in this temperature region.

Figure 10:
Magnetic Phase Diagram for $TmMnO_3$. The $P\underline{6}_3\underline{c}m$ phase becomes unstable above 10.5 Tesla. The phase boundary is are derived from anomalies of ε(T) and ε(H).

Figure 11:
ε(T) at high fields for $TmMnO_3$. Above 10.5 Tesla no anomaly can be detected. Different curves are vertically offset for clarity. H is applied along the c-axis.

Figure 12:
Magnetic Phase Diagram for $HoMnO_3$ extended to higher fields. For clarity, some subtle features of the low-field low-temperature phases are not shown.

Figure 13:
Magnetic field dependence of the c-axis length of $HoMnO_3$ at higher temperatures (8 K and 15 K).

Figure 14:
Low-temperature magnetic phase diagram of $HoMnO_3$ ($T<10$ K). Symmetries are assigned to some phases according to the results of second harmonic generation experiments [12].

Figure 15:
Field dependence of the c-axis of $HoMnO_3$ at low temperatures. The inset shows the magnetostrictive coefficient at 3 K with four subsequent peaks indicating successive phase transitions. The first (positive) and the second (negative) peak represent the entrance into and exit of the LT1 phase, respectively. The two higher field peaks are the $T_{3'}$ and $T_1$ phase boundaries.

Figure 16:
Low-temperature ε(H) along the c-axis of $HoMnO_3$. The phase boundary to the LT2 phase ($T_4$ in Fig.14) is marked by dashed arrows labeled $H_4$. At 1.4 K, the valley of c(H) between 0.5 and 2 Tesla represents the LT1 phase.

Fig. 1

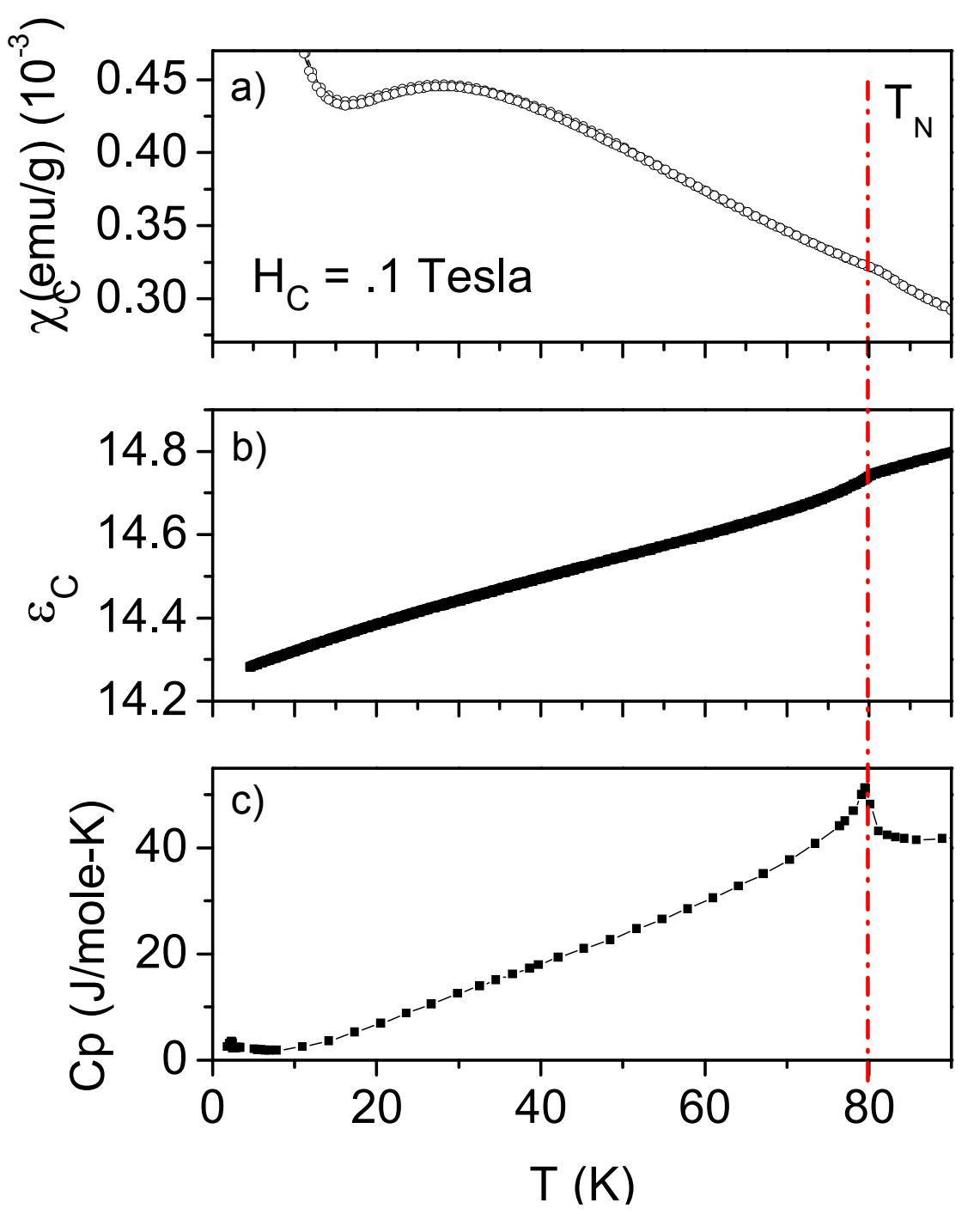

Fig. 2

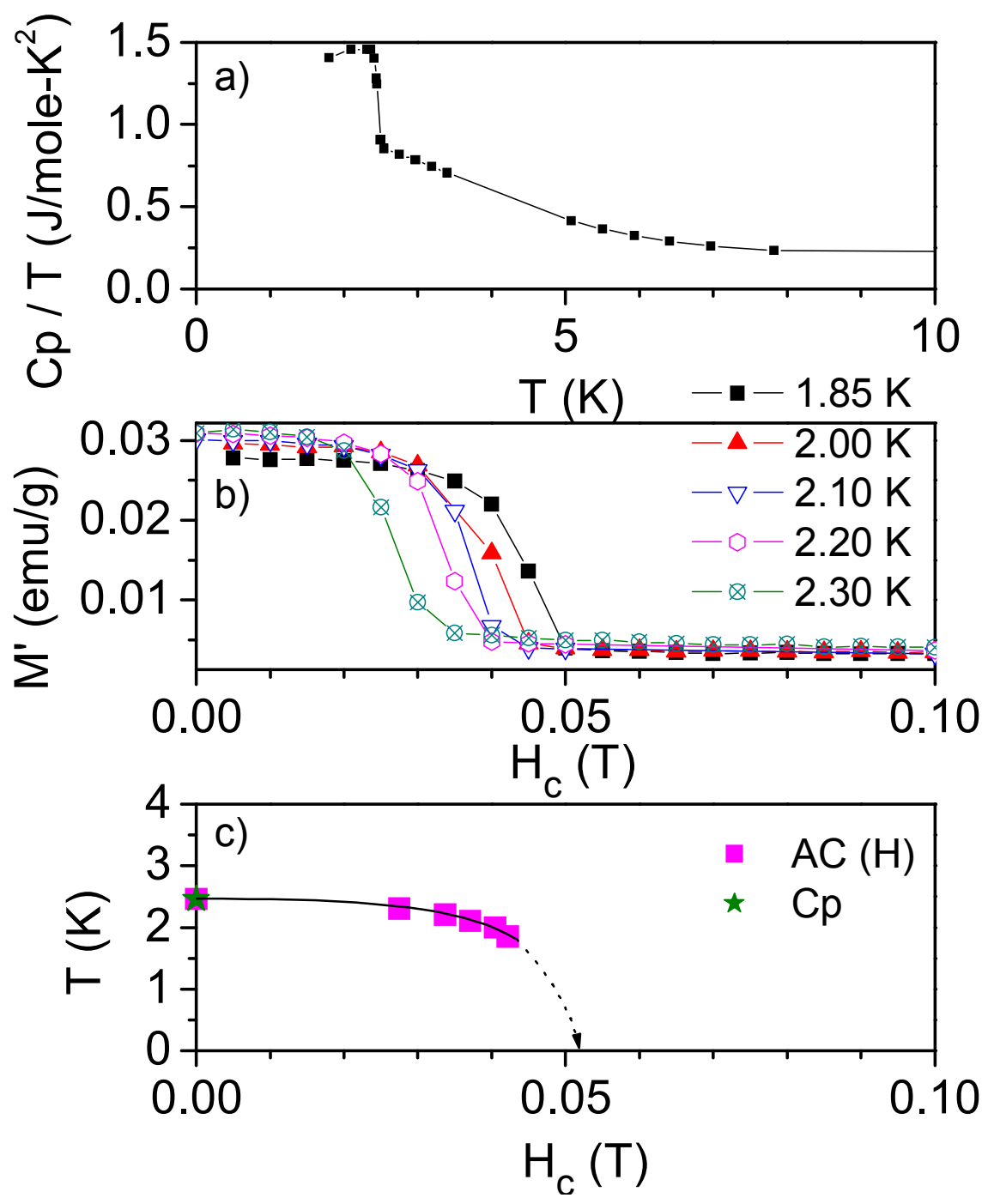

Fig 3

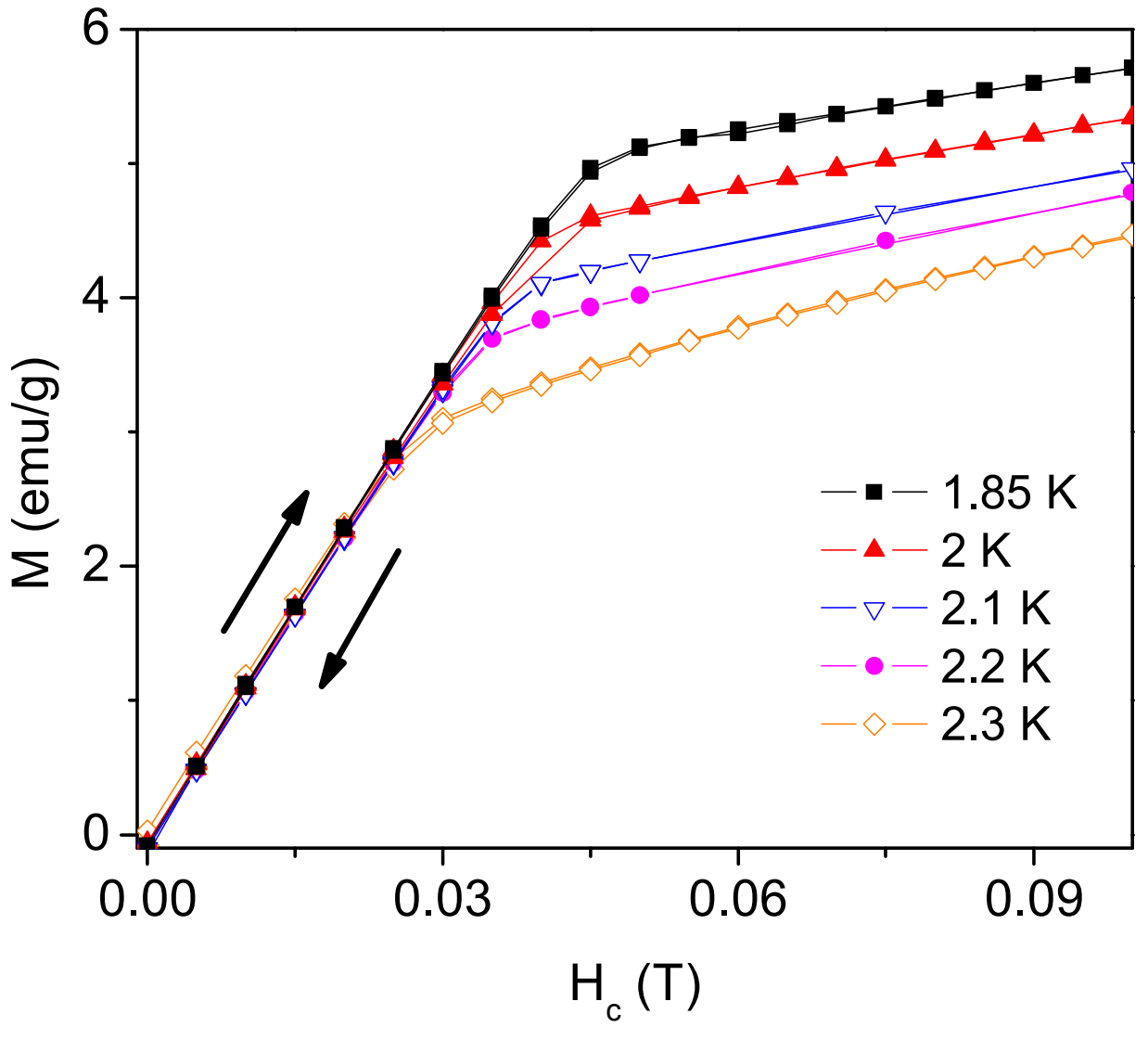

Fig. 4

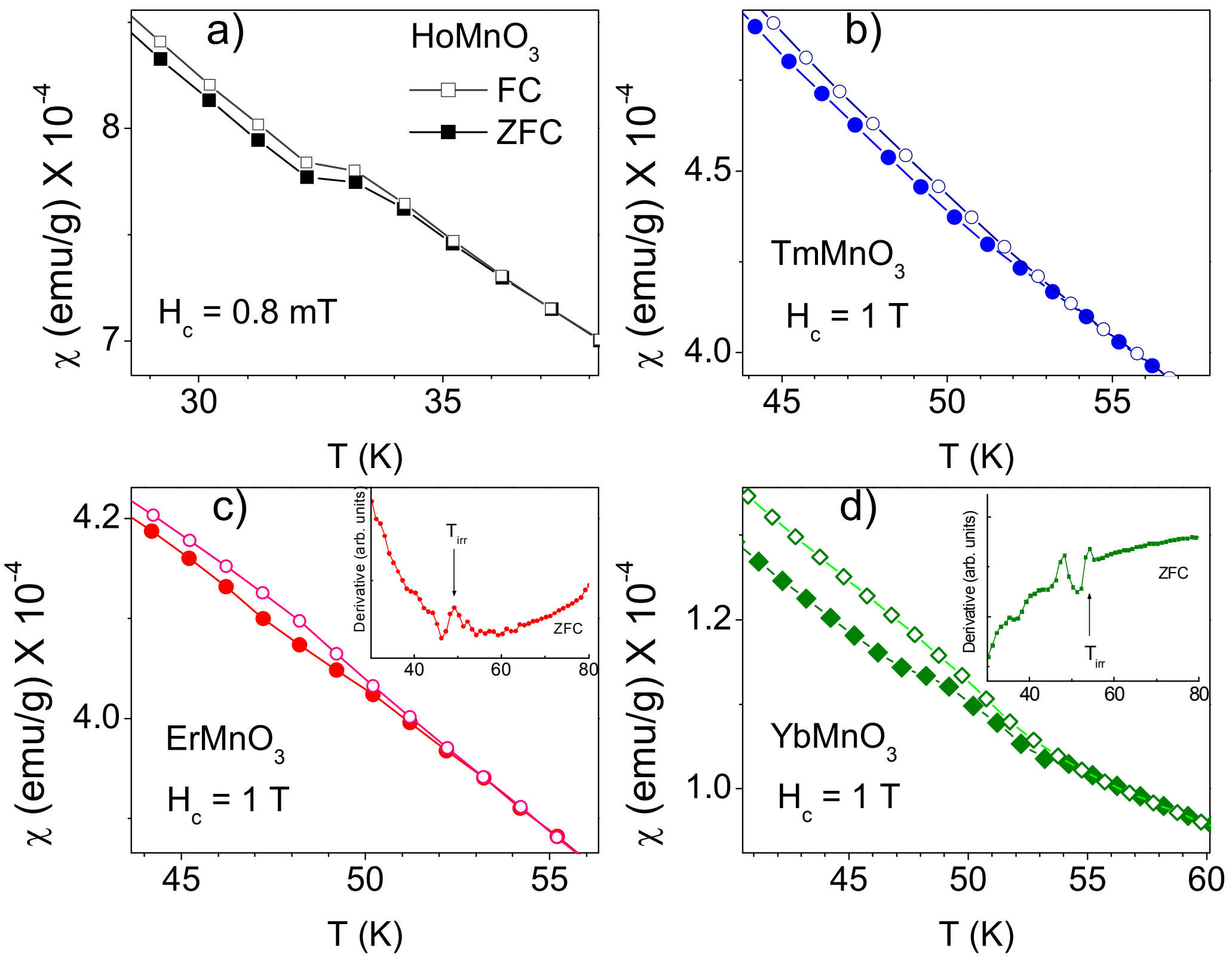

a)
HoMnO3
FC
ZFC
$H_c$ = 0.8 mT
b)
TmMnO3
$H_c$ = 1 T
c)
ErMnO3
$H_c$ = 1 T
Derivative (arb. units)
$T_{irr}$
ZFC
d)
YbMnO3
$H_c$ = 1 T
χ (emu/g) X $10^{-4}$
T (K)

Fig. 5

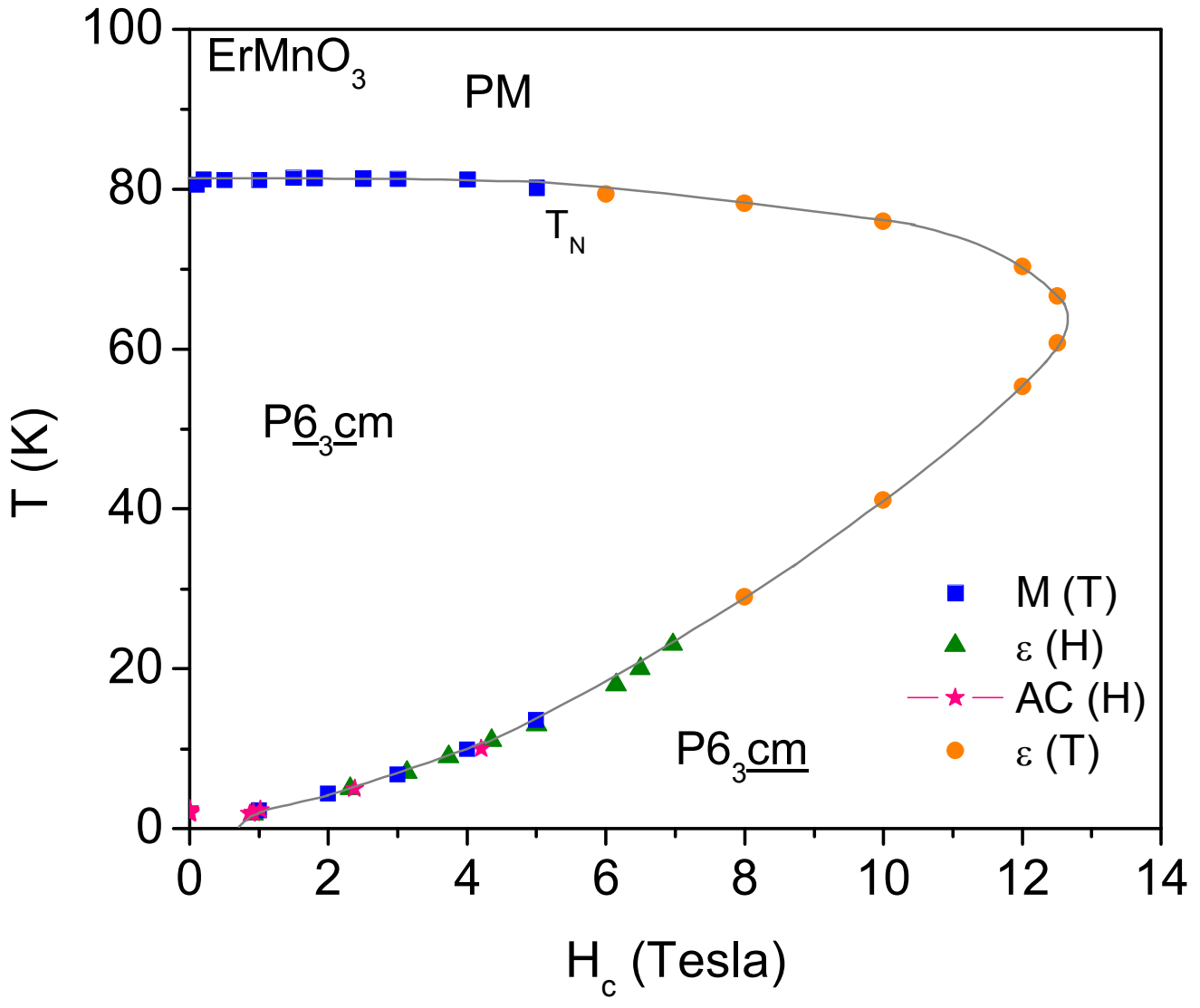

Fig. 6

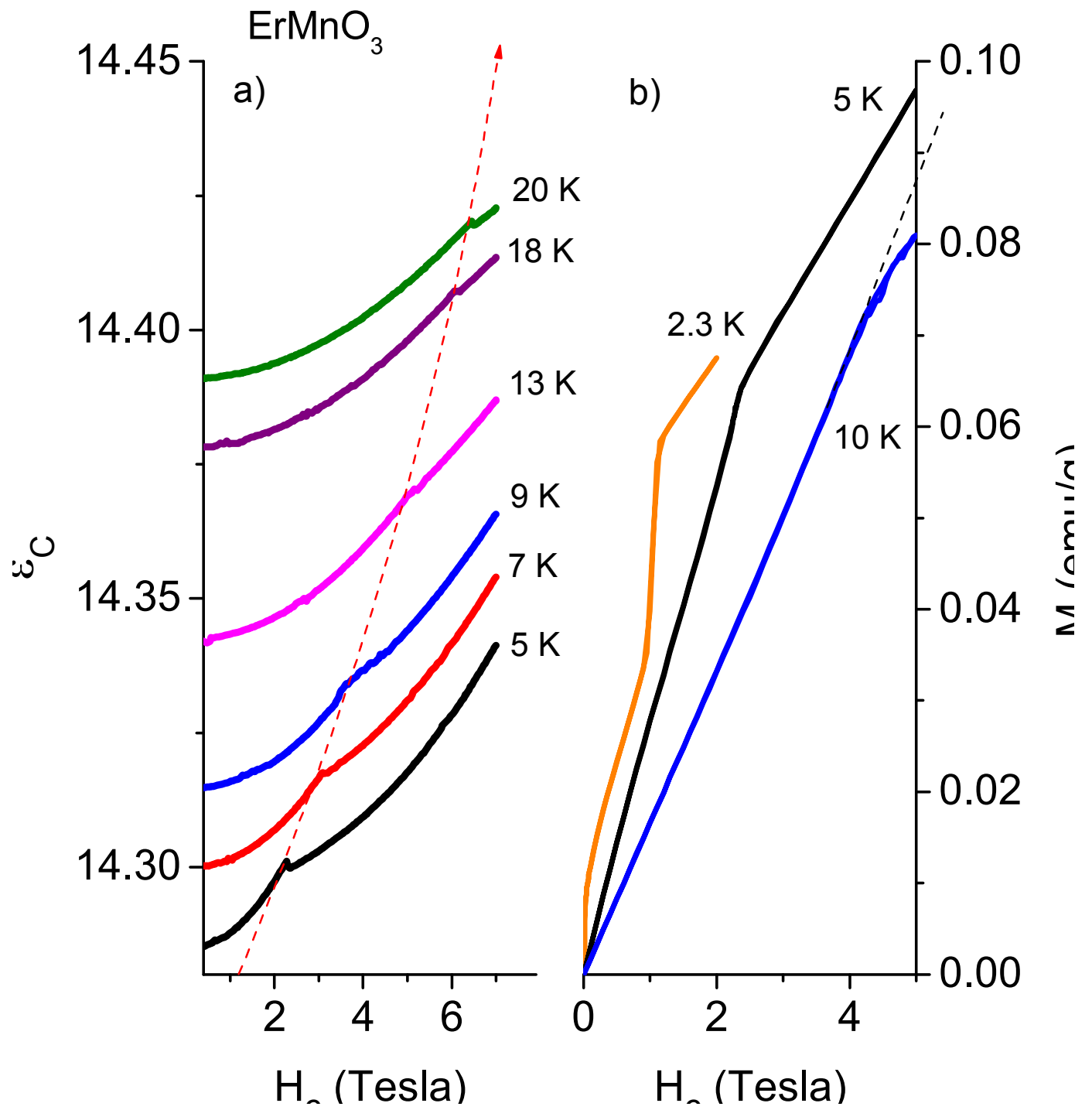

Fig. 7

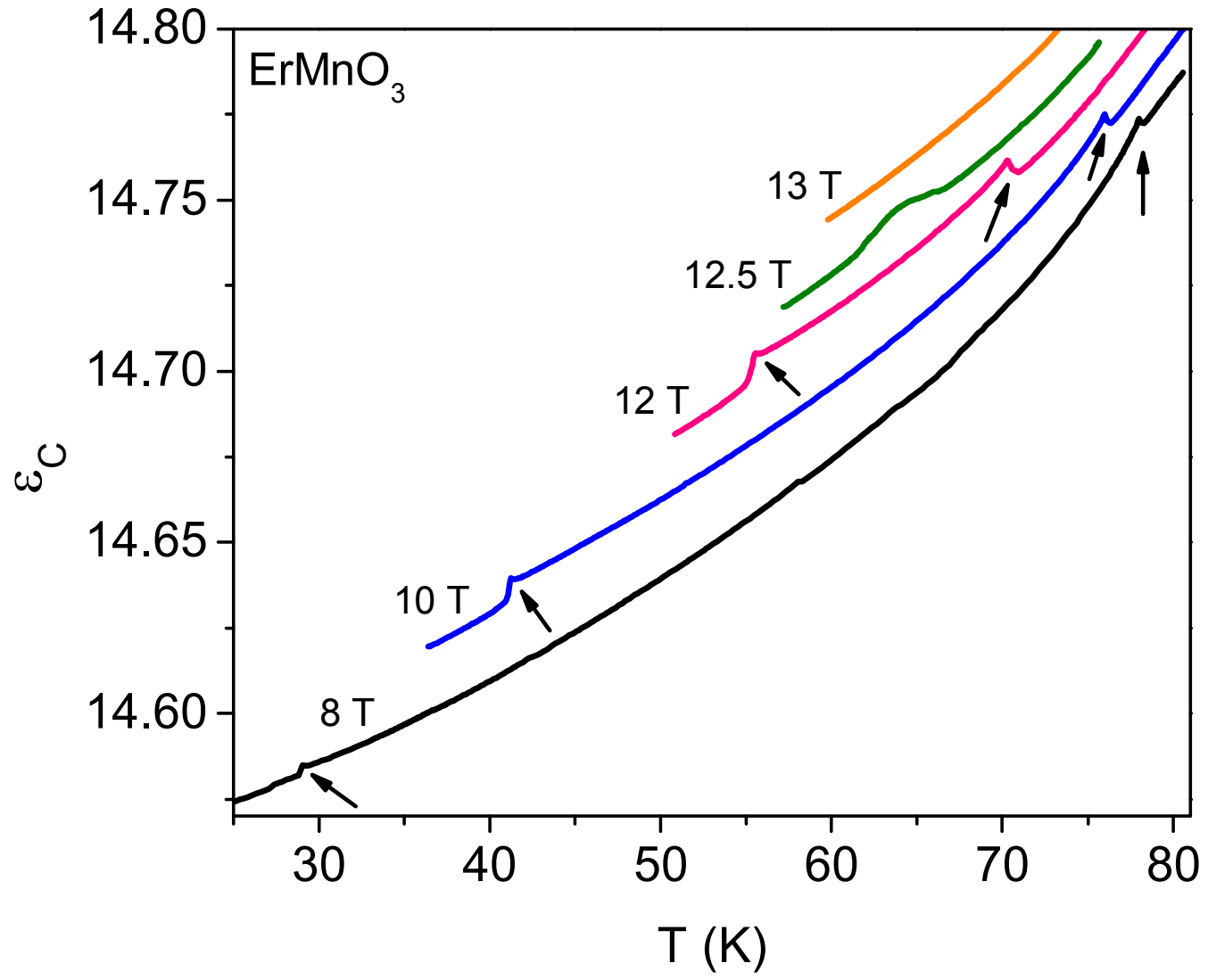

Fig. 8

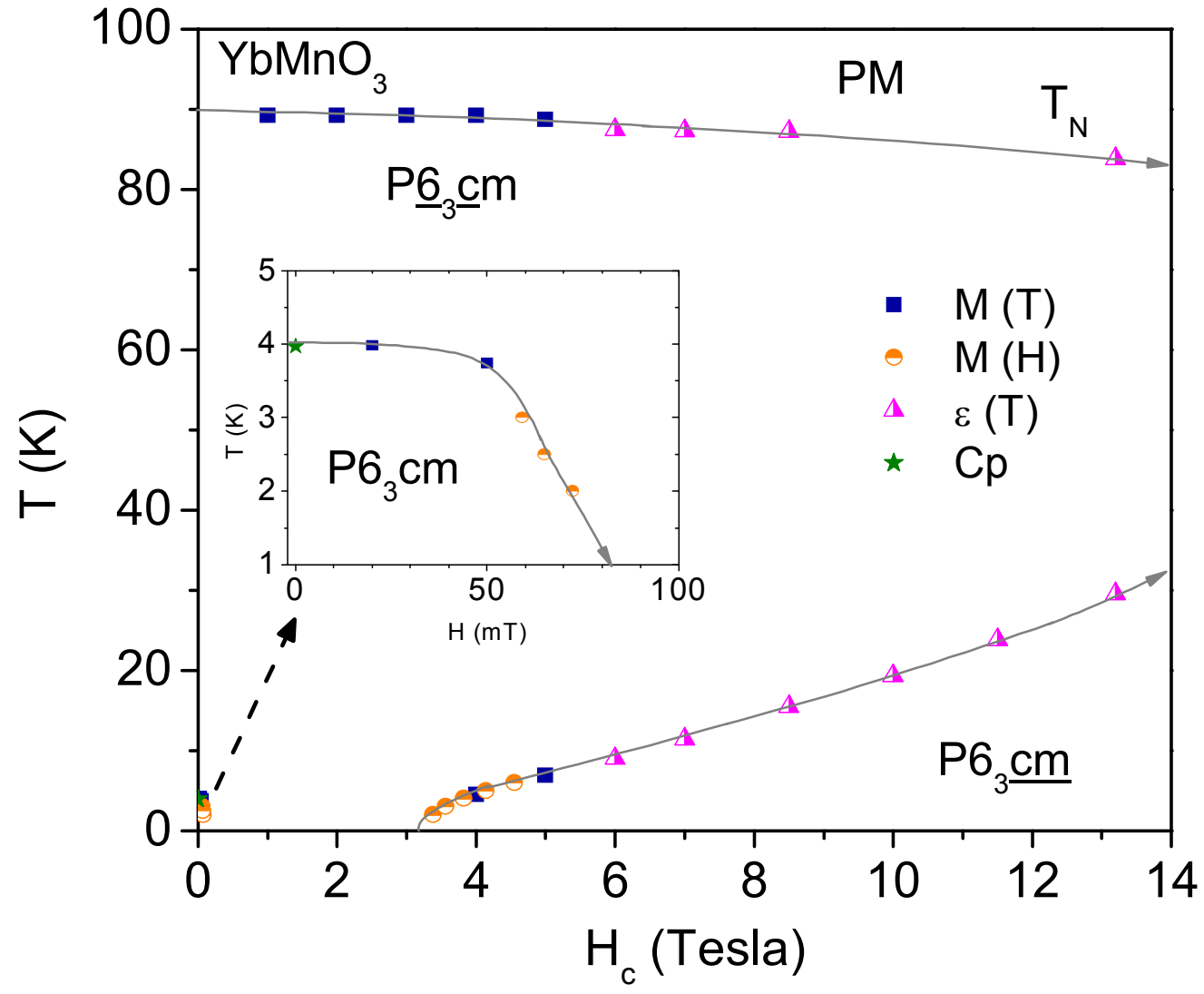

Fig. 9

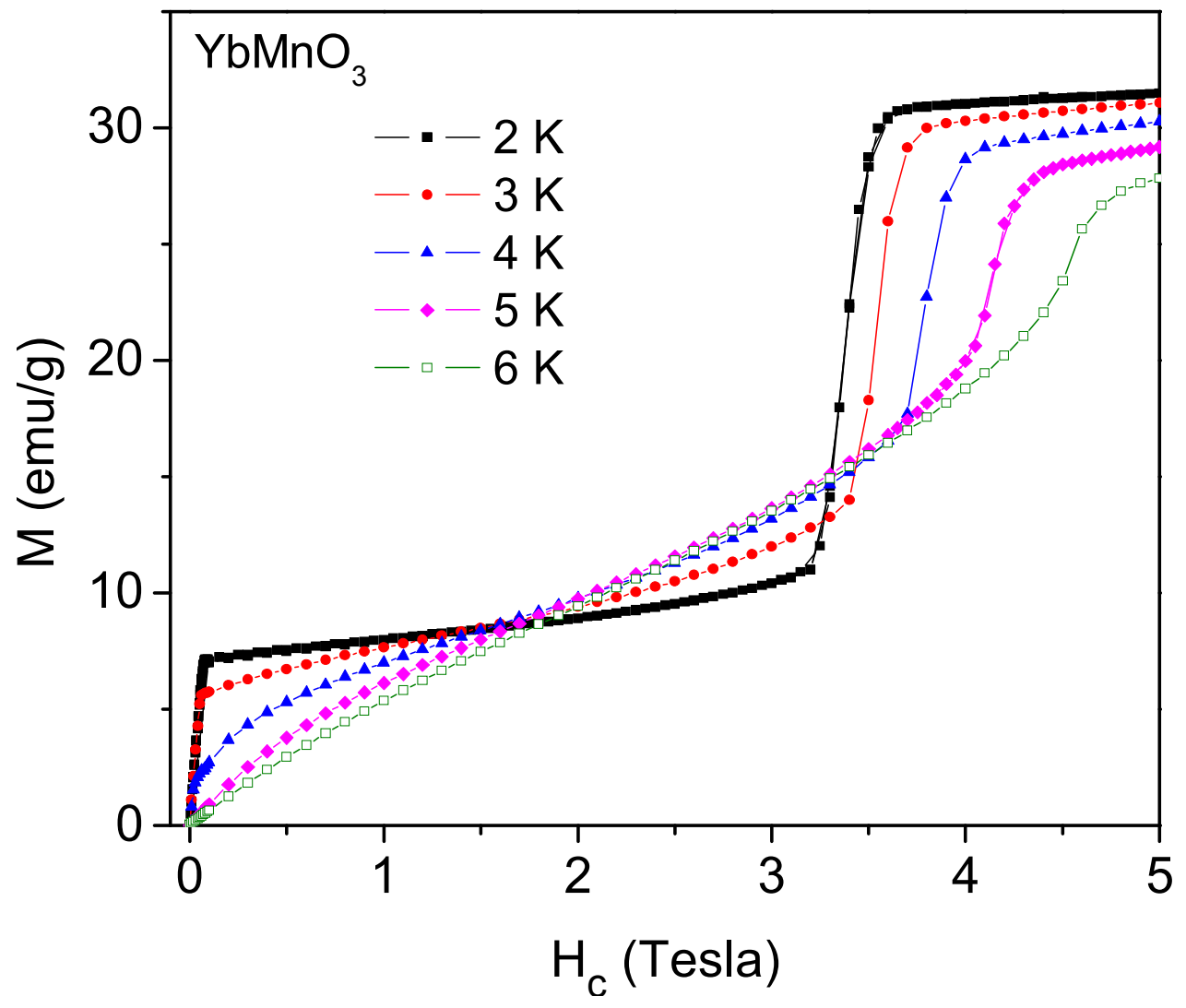

Fig. 10

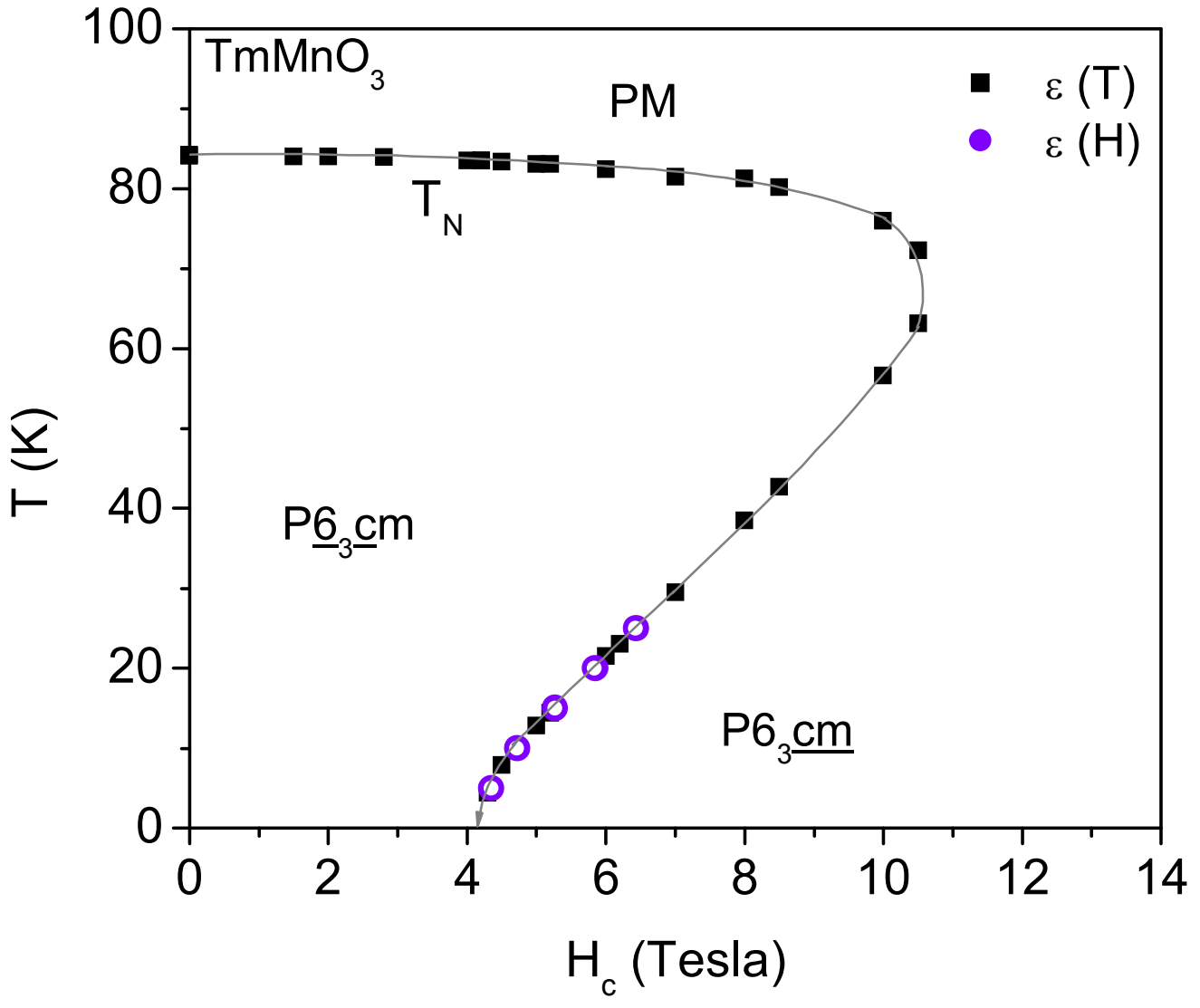

Fig. 11

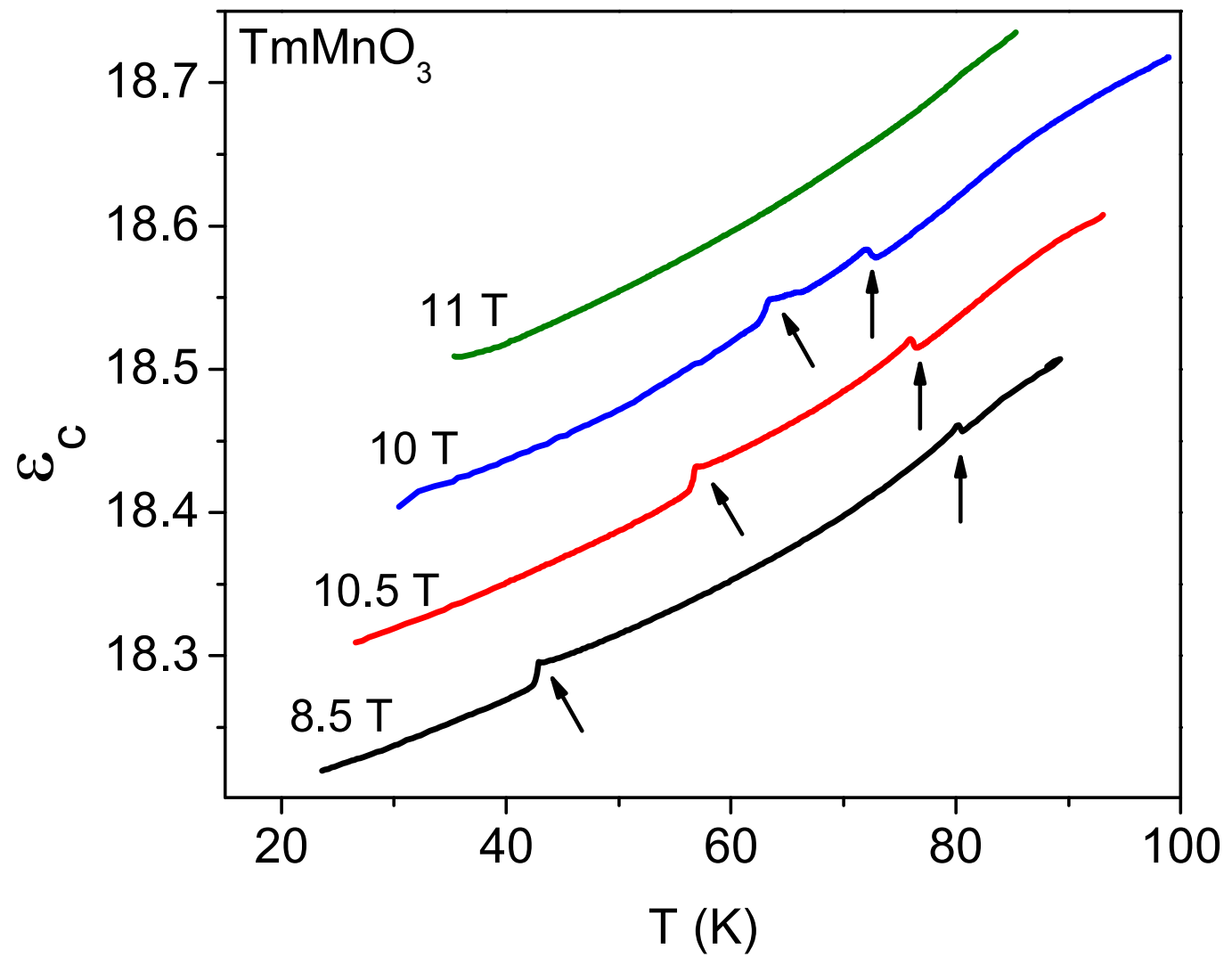

Fig. 12

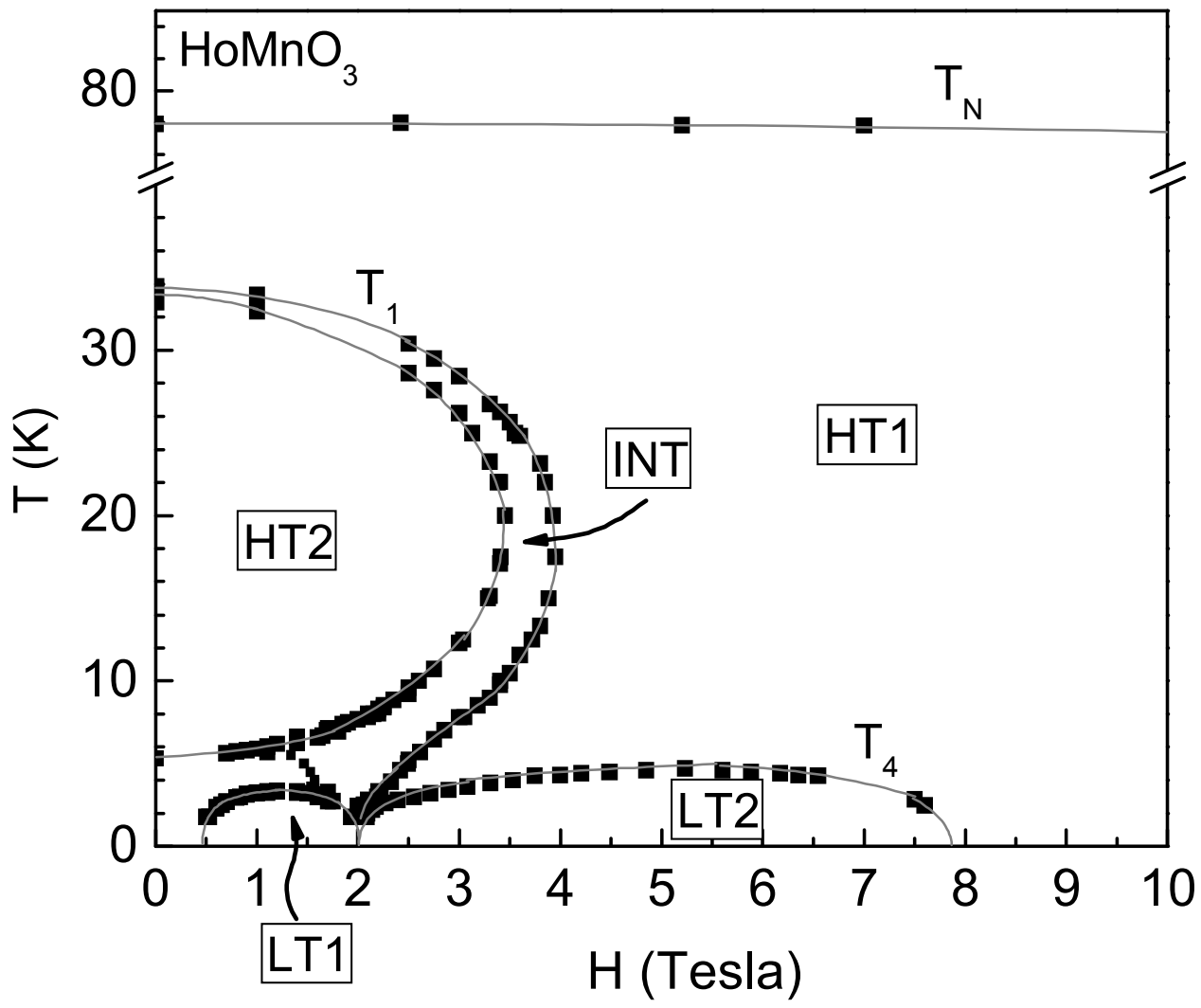

HoMnO3
TN
T1
T4
INT
HT1
HT2
LT1
LT2
80
30
20
10
0
T (K)
0 1 2 3 4 5 6 7 8 9 10
H (Tesla)

Fig. 13

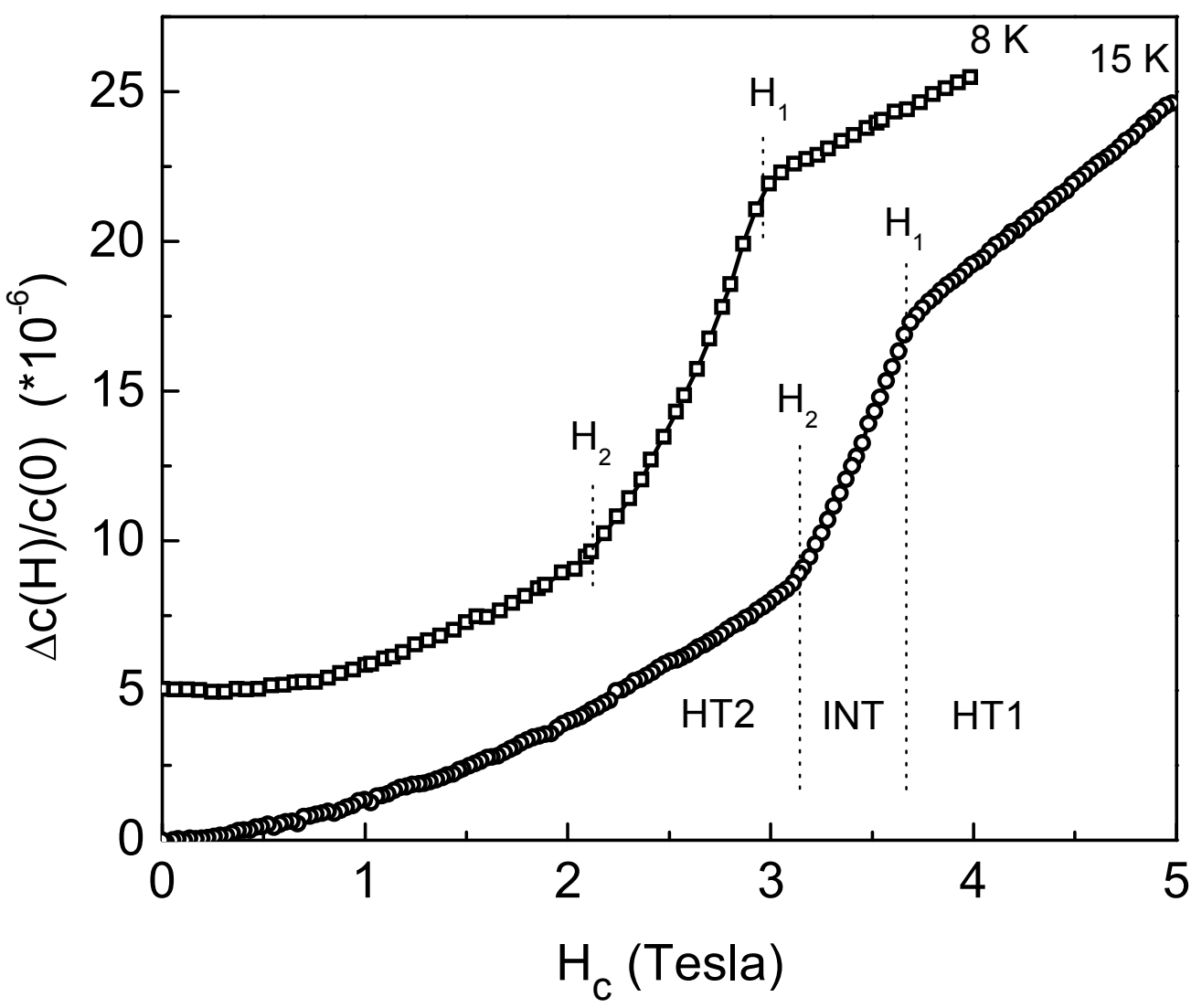

Fig. 14

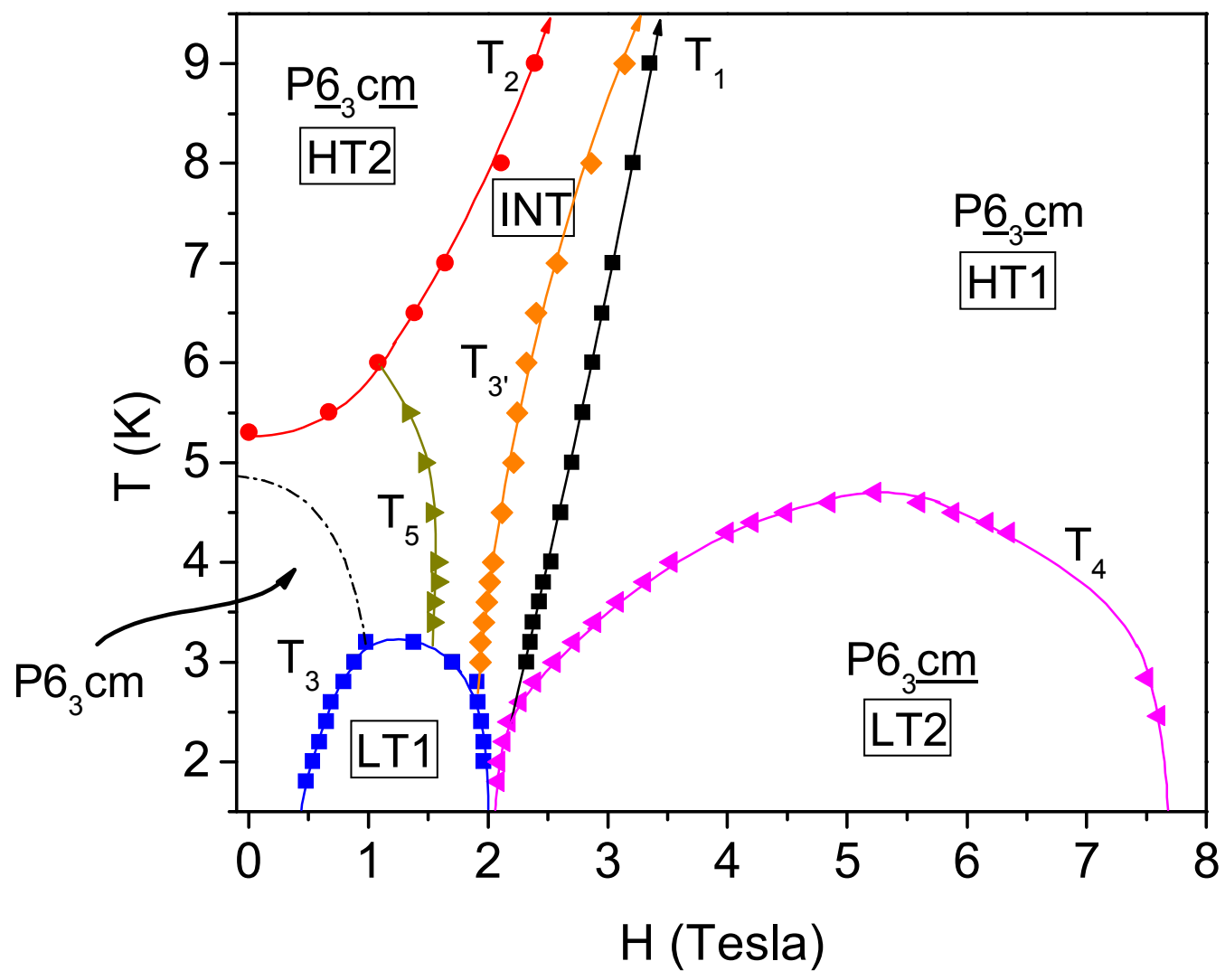

Fig. 15

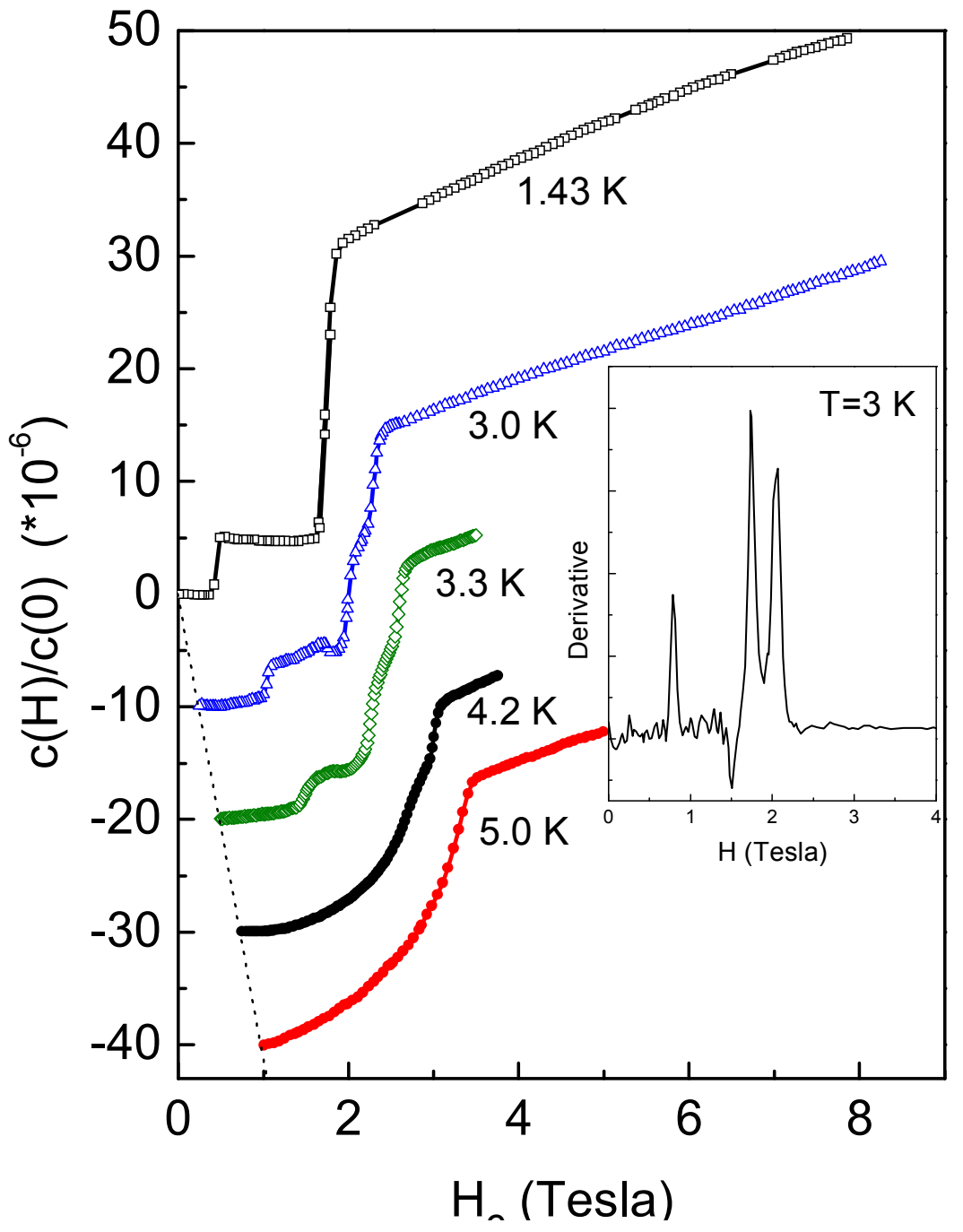

Fig. 16

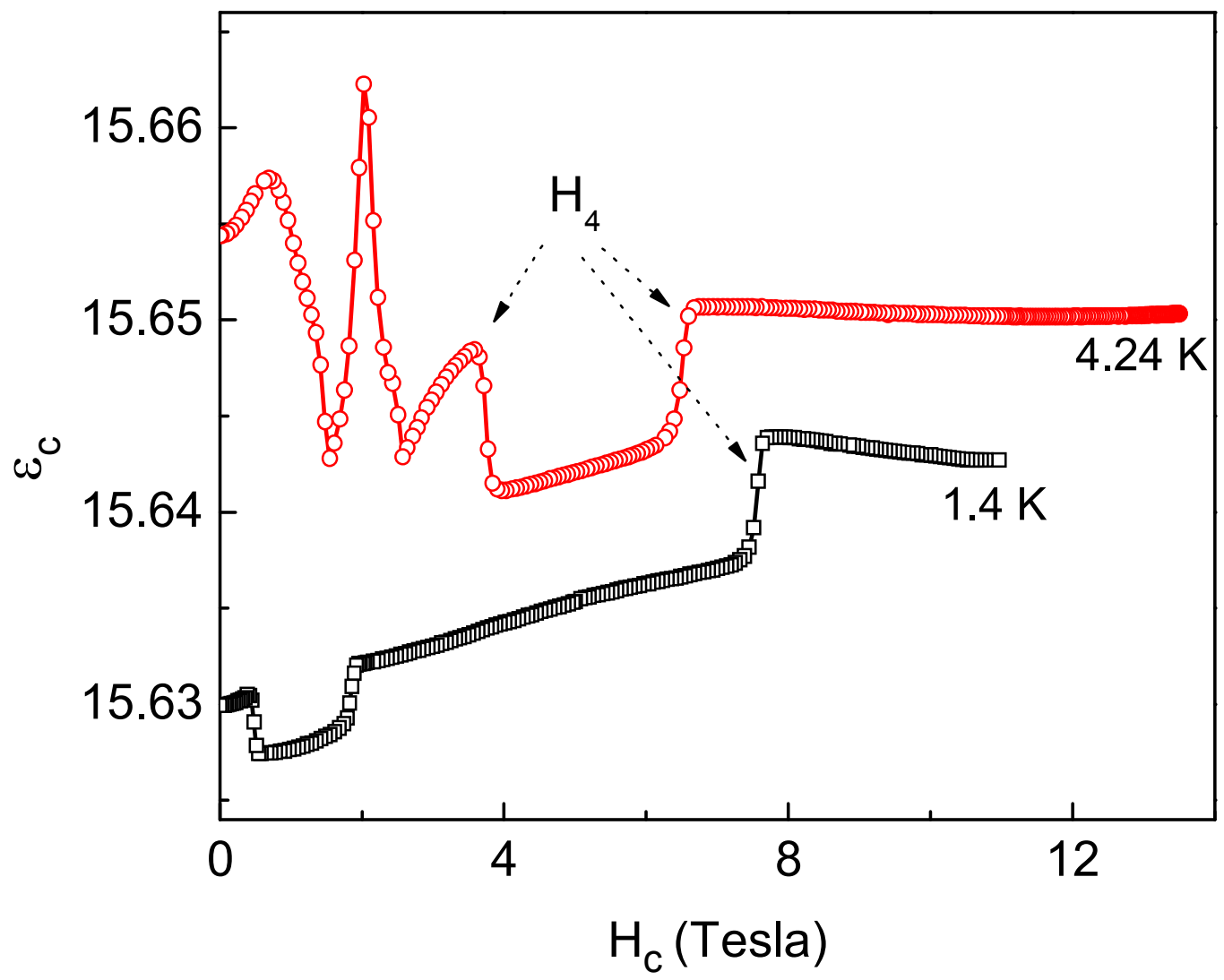

$\varepsilon_c$
H$_c$ (Tesla)
H$_4$
4.24 K
1.4 K